\documentclass[a4paper,11pt]{article}
\pdfoutput=1 

\usepackage{jcappub}
\usepackage{aas_macros}
\usepackage[cp1252]{inputenc}
\usepackage[T1]{fontenc} 
\usepackage{enumitem}

\title{\boldmath Non-parametric Lagrangian biasing from the insights of neural nets}

\author[a,b,1]{Xiaohan Wu,\note{Corresponding author.}}
\author[a]{Julian B.~Mu\~noz,}
\author[a]{Daniel J.~Eisenstein}

\affiliation[a]{Harvard-Smithsonian Center for Astrophysics, 60 Garden Street, Cambridge 02138, MA, USA}
\affiliation[b]{Canadian Institute for Theoretical Astrophysics, University of Toronto, Toronto, ON, Canada M5S 3H8}
\emailAdd{xwu@cita.utoronto.edu}
\emailAdd{julianmunoz@cfa.harvard.edu}
\emailAdd{deisenstein@cfa.harvard.edu}

\abstract{
We present a Lagrangian model of galaxy clustering bias in which we train a neural net using the local properties of the smoothed initial density field to predict the late-time mass-weighted halo field.
By fitting the mass-weighted halo field in the \textsc{AbacusSummit} simulations at $z=0.5$, we find that including three coarsely spaced smoothing scales gives the best recovery of the halo power spectrum.  Adding more smoothing scales may lead to 2-5\% underestimation of the large-scale power and can cause the neural net to overfit.
We find that the fitted halo-to-mass ratio can be well described by two directions in the original high-dimension feature space.  
Projecting the original features into these two principal components and re-training the neural net either reproduces the original training result, or outperforms it with a better match of the halo power spectrum.  The elements of the principal components are unlikely to be assigned physical meanings, partly owing to the features being highly correlated between different smoothing scales.
Our work illustrates a potential need to include multiple smoothing scales when studying galaxy bias, and this can be done easily with machine-learning methods that can take in high dimensional input feature space.
}

\begin{document}
\maketitle
\flushbottom

\section{Introduction}
\label{sec:intro}

Galaxy surveys that map the large-scale structure of the universe have been a powerful probe of the composition of our universe and how its structures form (\cite[for a recent review, see][]{desjacques18}). In the canonical picture, structure formation originates from the gravitational collapse of initially small perturbations in the matter density field, which eventually grow into dark-matter halos.  Galaxy formation then proceeds within these halos by gas accretion and cooling, producing galaxies that act as biased tracers of the underlying density field.  An improved understanding of where dark-matter halos sit in the underlying matter field is crucial for modeling the formation of the large-scale structure and therefore constraining cosmological parameters.
This, in turn, is key to understand the physics of inflation, and the dark sector of our universe.

Traditionally, the halo density is assumed to trace the underlying matter field in a polynomial form, with a series of bias parameters characterizing the relation \cite[e.g.][]{fry93, matsubara08, vlah16}.  This bias expansion approach has achieved much success in describing summary statistics such as the galaxy power spectrum and bispectrum \cite{chan12, baldauf12, saito14, abidi18, fujita20, modi20, kokron21} as well as halos at the field level \cite{schmittfull19, schmittfull20, modi19, barreira21}, and previous works have focused on evaluating the biases \cite{kaiser84, desjacques10, musso12, baldauf15, modi17, lazeyras16, lazeyras18, lazeyras19, lazeyras21}.
In Ref.~\cite{wu22} (hereafter Paper I), we developed a fully non-parametric framework to calculate the distribution of halos at a given redshift given the initial Lagrangian density field.  This goes beyond the traditional bias expansion approach and instead fits a function $f$ that characterizes how much halo mass should form given properties of the initial density field.  Specifically, we model the initial Lagrangian-space (pre-advection) halo overdensity $\delta_{\rm h}$ as
\begin{equation}
1+\delta_h = f(\delta, \nabla^2\delta, \mathcal{G}_2),
\end{equation}
where $\delta$ is the Lagrangian matter overdensity and $\mathcal{G}_2$ is the tidal operator, evaluated at each point in the initial Lagrangian space \cite{matsubara08, mcdonald09, assassi14, vlah16}.
We showed that for mass-weighted halos above a mass threshold, the shape of $f$ clearly deviates from a polynomial of $\delta$ and other quantities.
Therefore while the bias expansion approach has been successful and physically intuitive \cite[from the peak-background split argument,][]{bardeen86, mo96, sheth99}, it can yield an unphysical relation between the galaxy and matter fields such as a non-positive-definite $f$ and enhanced biases for underdense regions.
We demonstrated that our fitted $f$ can recover the late-time halo power spectrum to sub-percent level at wavenumbers $k<0.1\ h$~Mpc$^{-1}$

In this work we expand upon our previous Paper I and train a neural network (NN) to obtain $f$ as a function of particle features associated with various smoothing scales.  Our previous formalism involves dividing the feature space into a finite number of bins and solving for a piece-wise constant $f$ with least-squares fitting.  This limited us to using at most two features (such as $\delta$ and $\nabla^2\delta$) owing to memory restrictions.  A NN, however, allows $f$ to be a continuous function of a large number of features.  We show that with $\delta, \nabla^2\delta, \mathcal{G}_2$ associated with three smoothing scales appropriately chosen, the $f$ function predicted by the NN is able to recover the halo power spectrum to within 1-2\% at $k\lesssim 0.1\ h$~Mpc$^{-1}$ despite being trained at the field level.  We find that for mass-weighted halos, $f$ can be well described by two orthogonal directions in the original feature space.  Projecting the original $\delta, \nabla^2\delta, \mathcal{G}_2$ into these two directions and re-training the NN reproduces the original training results of the halo power spectrum at least as well.
While our work is closely related to Refs.~\cite{luciesmith18, luciesmith19, luciesmith20} who also studied halo formation and large-scale structure using machine learning methods, we focus on predicting the entire halo field instead of halo masses as done in these works.

This paper proceeds as follows.  Section~\ref{sec:methods} presents our setup of the NN and the simulations.  Section~\ref{sec:results_Rf} discusses results of using different combinations of smoothing scales.  Section~\ref{sec:results_PC} examines the structure of $f$ using the principal components of $\nabla f$.  We briefly explore modeling a thin mass range instead of a mass threshold in Section~\ref{sec:thin_mass_range}, and discuss future directions Section~\ref{sec:conclusions}.

\section{Methods}
\label{sec:methods}

\subsection{Neural Net}

We aim to train a NN to predict the $f$ weight that a particle should carry given its input features, so that such an $f$ function best recovers the real-space halo field $\delta^{\rm true}_h$.
To obtain the particle features, we first compute the smoothed overdensity field $\delta$, its Laplacian $\nabla^2\delta$, and the corresponding tidal shear $\mathcal{G}_2$ using the initial condition of a simulation given a smoothing scale $R_f$.  These fields are normalized to have standard deviation of 1.
We then assign each particle in a simulation a certain set of $\delta, \nabla^2\delta, \mathcal{G}_2$ values according to its nearest grid point in the initial condition.  In the presence of multiple smoothing scales $R_{f1}, R_{f2}, ...$, a particle carries a vector of features
\begin{equation}
\theta_{\rm orig} = (\delta_1, \nabla^2\delta_1, \mathcal{G}_{2,1}, \delta_2, \nabla^2\delta_2, \mathcal{G}_{2,2}, ...)^T,
\end{equation}
where the subscripts denote the indices of the smoothing scales.
These features are not independent of each other.  Specifically, the $\delta$'s and $\nabla^2\delta$'s with different smoothing scales are highly correlated with each other.  The $\mathcal{G}_2$'s are uncorrelated with $\delta$'s and $\nabla^2\delta$'s since these are quadratic terms, but are highly correlated among themselves.

To expose the independent degrees of freedom to the NN, we opt to 
orthogonalize the original features.  We thus calculate the covariance matrix of the features
\begin{equation}
\Sigma_{ij} = (\theta_{{\rm orig},i}, ...) \cdot (\theta_{{\rm orig},j}, ...)^T,
\end{equation}
where the subscripts $i,j$ denote the indices of particles, and the size of the covariance matrix is the number of original features squared.
We then obtain the eigenvalues $\Lambda$ and eigenvectors $V$ of $\Sigma$, where $\Lambda$ is a diagonal matrix with the eigenvalues as the diagonal elements, and the columns of the matrix $V$ are the corresponding eigenvectors.  Defining a matrix $\Lambda^{-1/2}$, which is diagonal and the elements are the inverse of the square root of $\Lambda$, the orthogonalized and normalized features are
\begin{equation}
\theta_{\rm orth} = \Lambda^{-1/2} V^T \theta_{\rm orig}.
\label{eq:feature_orthogonalization}
\end{equation}

Since the original features are highly correlated among each other, the range of the square root of their eigenvalues can span three or four magnitudes.  It is only desirable to keep the transformed features with larger eigenvalues.  Intuitively, the $\nabla^2\delta$'s can be written as the derivative of the smoothed $\delta$'s with respect to the smoothing scale, so can be approximated with a linear combination of $\delta$'s with adjacent smoothing scales.  For a number $n_{R_f}$ of fine-spaced smoothing scales, we thus only need the $n_{R_f}$ $\delta$'s and one $\nabla^2\delta$ with the smallest smoothing scale to approximate all the $2n_{R_f}$ $\delta$'s and $\nabla^2\delta$'s.  This motivates us to keep the $n+1$ eigenvectors with the largest eigenvalues calculated using only $\delta$'s and $\nabla^2\delta$'s.\footnote{We have verified that for small separation of the smoothing scales, inputting $n_{R_f}$ $\delta$'s and only the $\nabla^2\delta$ of the smallest smoothing scale yields the same training results as inputting $2n_{R_f}$ $\delta$'s and $\nabla^2\delta$'s and keeping the largest $2n_{R_f}+1$ eigenvectors.  This no longer holds with large separations of the smoothing scales.}  We keep all of the $n_{R_f}$ $\mathcal{G}_2$'s, since in the case of coarsely separated $R_f$'s the associated $\mathcal{G}_2$'s are not highly degenerate.\footnote{We find that keeping $2n_{R_f}+1$ features instead of choosing the features based on a cut-off of the ratios of the eigenvalues yields more stable training results.  Specifically, if the eigenvectors and eigenvalues have small changes when calculated from a different simulation box, the resulting $f$ appears to be robust.}  This leaves us with a total of $2n_{R_f}+1$ orthogonalized and normalized features that should be input into the NN, which we represent by a projection matrix $P$
\begin{equation}
\theta_{\rm input} = P \theta_{\rm orth} = P \Lambda^{-1/2} V^T \theta_{\rm orig}.
\label{eq:feature_transformation_final}
\end{equation}
Here $P$ is a diagonal matrix of size $n^2$, with $2n_{R_f}+1$ of the diagonal elements being 1 and the others being 0, which projects out the unused $\theta_{\rm orth}$.

For each particle, given its orthogonalized and normalized features $\theta$, a NN predicts an $f$ value that it should carry.  We then sum up the $f$ values of particles in each cell to obtain the model halo field:
\begin{equation}
1+\delta^{\rm model}_{h,j} = \frac{N_{\rm cell}}{N_{\rm part}} \sum_{i\in {\rm cell}_j} f(\theta_{{\rm input},i})
\end{equation} 
where $j$ denotes the index of a cell.  Here the sum of $f$ should be multiplied by the size of the halo grid $N_{\rm cell}$ divided by the total number of particles $N_{\rm part}$ to ensure that the mean of the halo overdensity is 0.  The loss function is defined as
\begin{equation}
L = \sum_j (\delta^{\rm model}_{h,j} - \delta^{\rm true}_{h,j})^2.
\end{equation}
The summation is carried over each batch in the training, where a batch is a number of grid cells with the particles in them.

In addition to the squared loss, the predicted $f$'s should be non-negative and satisfy the integral constraint $\sum_i f_i / N_{\rm part} = 1$ owing to mass conservation, where $i$ denotes the particle indices.  To enforce the non-negativity constraint, we make the NN predict $y=\log_{10}(f)$ instead of $f$, and add an additional penalty $(y+5)^{4}$ for each particle to prevent $y$ from going too negative and thus yielding a vanishing gradient.
To implement the integral constraint, we choose the batch size so that each batch, which is a collection of cells, contains about $N_{\rm part,batch} = 10^5$ particles in total.  The batches are randomly initialized before the training and kept unchanged during the training.  For each batch, in addition to the squared loss and the non-negativity loss, we add another loss $(\sum_i f_i / N_{\rm part,batch} - 1)^2 / \epsilon^2$, where $\epsilon$ is the tolerance.  During the first epoch of training, we do not include the integral constraint, and the NN recovers the integral constraint at a few percent level.  Starting from the second epoch, we set $\epsilon$ as a function of the number of epochs $\max(10^{(-1-3\log_{10}({\rm epoch}))}, 10^{-4})$, so that it gradually decreases.  This ensures that NN recovers the integral constraint at sub-percent level.

\begin{figure}
\centering
\includegraphics[width=\linewidth]{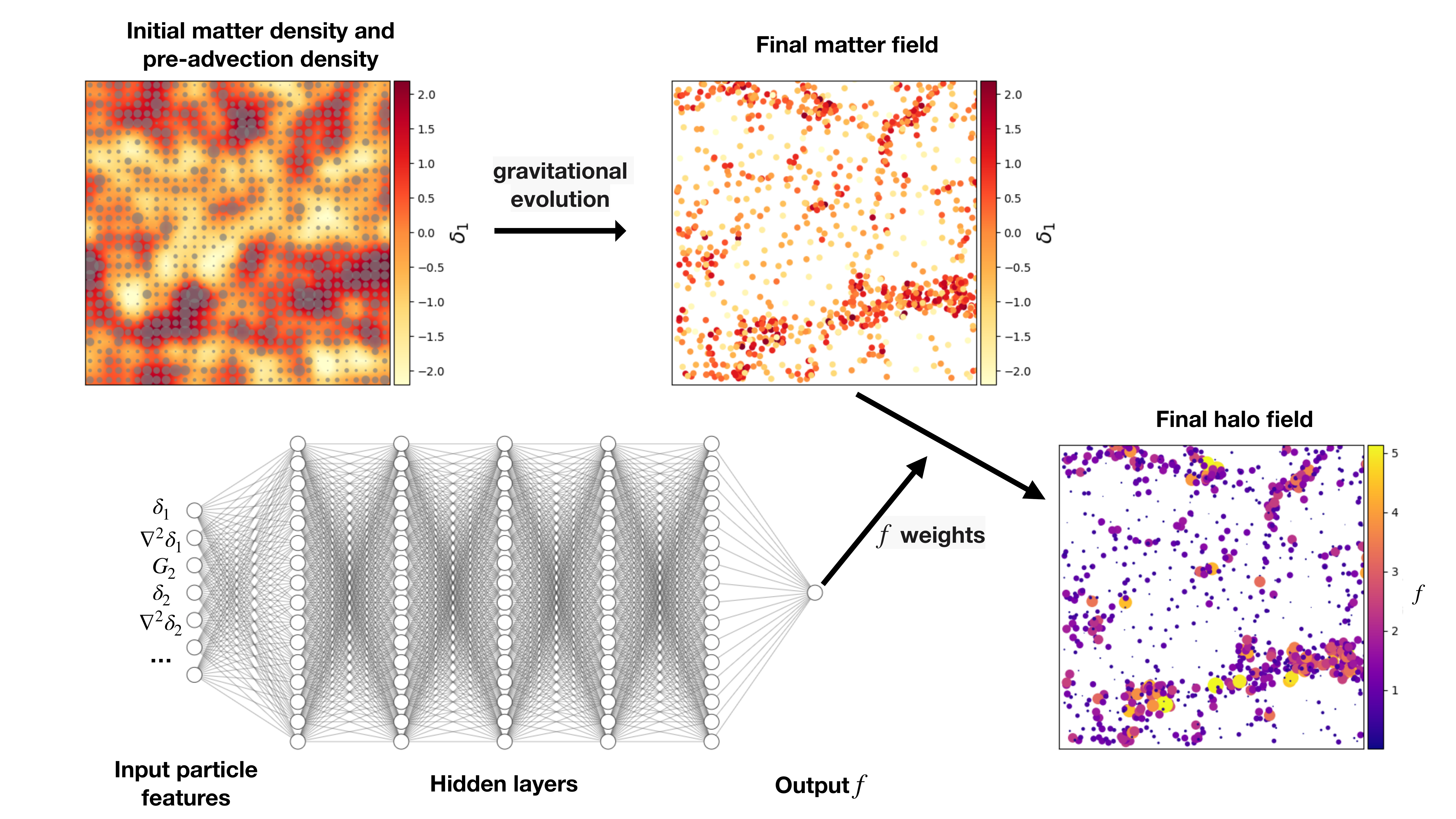}
\caption{A schematic illustration of our training process.  Top left panel shows the initial matter density field, illustrated by colors.  The overlying gray circles represent the $f$ weights that the particles should carry, with the size of the circles showing the amplitude of $f$.  The $f$ weights are predicted by the NN shown at the bottom left, by inputting the initial $\delta, \nabla^2\delta, \mathcal{G}_2$ at various smoothing scales.  The top right panel illustrates the final matter field, where the particles shown by circles have been moved by gravity.  The bottom right panel presents the final halo field, where the circles follow the locations of the particles and the colors and sizes of the circles represent the amplitude of $f$.  We calculate the loss based on the residual in modeling the final halo field.}
\label{fig:illustration}
\end{figure}

We stress that our NN makes predictions for particles, but our loss function is defined on the halo grid.  The backward propagation is handled internally by {\tt pytorch} as long as the loss function is implemented correctly.
Figure~\ref{fig:illustration} gives a schematic illustration of the training process, where the particles carry their corresponding $f$ weights predicted from the NN (gray circles in the top left panel) given at the initial time to their final locations, forming the final halo field (bottom right panel).

\subsection{Simulations and training}

We use the {\it AbacusSummit} simulations \cite{maksimova21} in this work to perform the training, which are run with the {\tt Abacus} N-body simulation code \cite{garrison18, garrison19, garrison21, metchnik09}.  These simulations were designed to meet and exceed the currently stated cosmological simulation requirements of the Dark Energy Spectroscopic Instrument (DESI) survey \cite{desi1}.  We utilize the small box simulations which have a box size of $500\ h^{-1}$~Mpc and $1728^3$ particles, using the Planck2018 standard cosmology \cite{planck18}: $\Omega_{\rm m}=0.14237, h=0.6736, \sigma_8=0.807952$.  This gives a particle mass of $2\times10^9\ h^{-1}\ M_\odot$.  Halos are identified on the fly with the {\tt CompaSO} halo finder which uses a hybrid FoF-SO algorithm \cite{hadzhiyska22}.  We focus on fitting the mass-weighted halo field with halo masses $M>3\times10^{11}\ M_\odot$ at $z=0.5$, corresponding to $>150$ particles.

The initial conditions were generated at $z=99$ using the method proposed in \cite{garrison16}.  To obtain the $(\delta, \nabla^2\delta, \mathcal{G}_2)$ values associated with a particle, we interpolated the initial density field onto $576^3$ grids and calculated the smoothed $(\delta, \nabla^2\delta, \mathcal{G}_2)$ values on each grid point given a smoothing scale $R_f$.  We then assign $(\delta, \nabla^2\delta, \mathcal{G}_2)$ values to each particle by looking for the nearest grid point to the particle's location in the initial space.  These are then input into the NN for the training.

To train a NN, we produce halo grids by summing the $f$ values of particles inside each cell.  We note that this calculation of $\delta^{\rm model}_h$ corresponds to using nearest neighbor interpolation (NNB), so the true halo field $\delta^{\rm true}_h$ is computed with NNB as well.  We use $100^3$ grids, which gives a cell size of $5\ h^{-1}$~Mpc.
To reduce the computational cost, for each grid cell, we only sample 10\% of the particles in it to perform the training.  This fraction guarantees that for our grid size of $5\ h^{-1}$~Mpc, there is at least one particle in each cell.  The summed $f$'s should then be scaled up by a factor of 10 to correctly obtain $\delta^{\rm model}_h$.

We implement a NN with 5 hidden layers, each with 64 neurons.  We use the GeLU activation function for all layers and the Adam optimizer for gradient descent.  We adopt an initial learning rate of $0.02$, which decays as $0.9^{\rm epoch}$.  This exponential decay ensures that the loss becomes rather flat after 20 epochs of training, and after 50 epochs of training the model predicted halo power spectrum converges as well.  We chose these hyperparameters so that in the case of one smoothing scale, the model predicted halo power spectrum changes at $<1-2\%$ level under different initializations of the NN and different sub-sampling of the particles.  Utilizing even larger NNs does not improve the performance on the halo power spectrum any further.

Each training process takes one simulation as the training set and other simulations as the validation set.
We train the NN five times and calculate the average $f$, each time with a different sub-sampling of the particles and initialization of the NN weights.  During a training process, we do not change the sub-sampled particles.
To examine the halo power spectrum, we apply the trained $f$ function onto all particles and create the halo field on a $256^3$ grid with Cloud-In-Cell interpolation.  We choose to use this finer grid when computing the halo power spectrum to avoid aliasing effects \cite{jing05}.  We will assess the range of scatter in power owing to the changes of $f$ in different times of training.

\section{Exploring different combinations of smoothing scales}
\label{sec:results_Rf}

\subsection{Comparison to least squares}

As a first crosscheck, we begin by comparing the NN results to those of the least-squares formalism we developed in Paper I.  In Paper I, we only include the $\delta$ and $\nabla^2\delta$ values of one smoothing scale for the particles, and divided the $\delta$-$\nabla^2\delta$ plane into bins according to the percentiles of $\nabla^2\delta$ at each $\delta$ value.  We then obtained the least-squares solution to $f$ in each bin by minimizing the mean squared error of the halo field, with the constraints that $f$ is non-negative and should integrate to 1.

Using the mass-weighted halo field with $M>3\times10^{11}\ M_\odot$ of one training simulation, we calculate the least-squares solution of $f$ with $R_f=2.83\ h^{-1}$~Mpc.  To make fair comparison, we train a NN with the $\delta$ and $\nabla^2\delta$ of this $R_f$.  Our $R_f$ choice is based on the mass-weighted mean mass of $M>3\times10^{11}\ h^{-1}\ M_\odot$ being $2.4\times10^{13}\ h^{-1}\ M_\odot$, corresponding to a Gaussian filter with $R_f=2.6\ h^{-1}$~Mpc.  Since we train the NN five different times each with a different sub-sampling of particles, we also sub-sample 10\% of the particles five times to obtain the least-squares solution.  We then compute the average $f$ solution and apply it on one validation simulation to calculate the halo power spectrum.

\begin{figure}
\centering
\includegraphics[width=\linewidth]{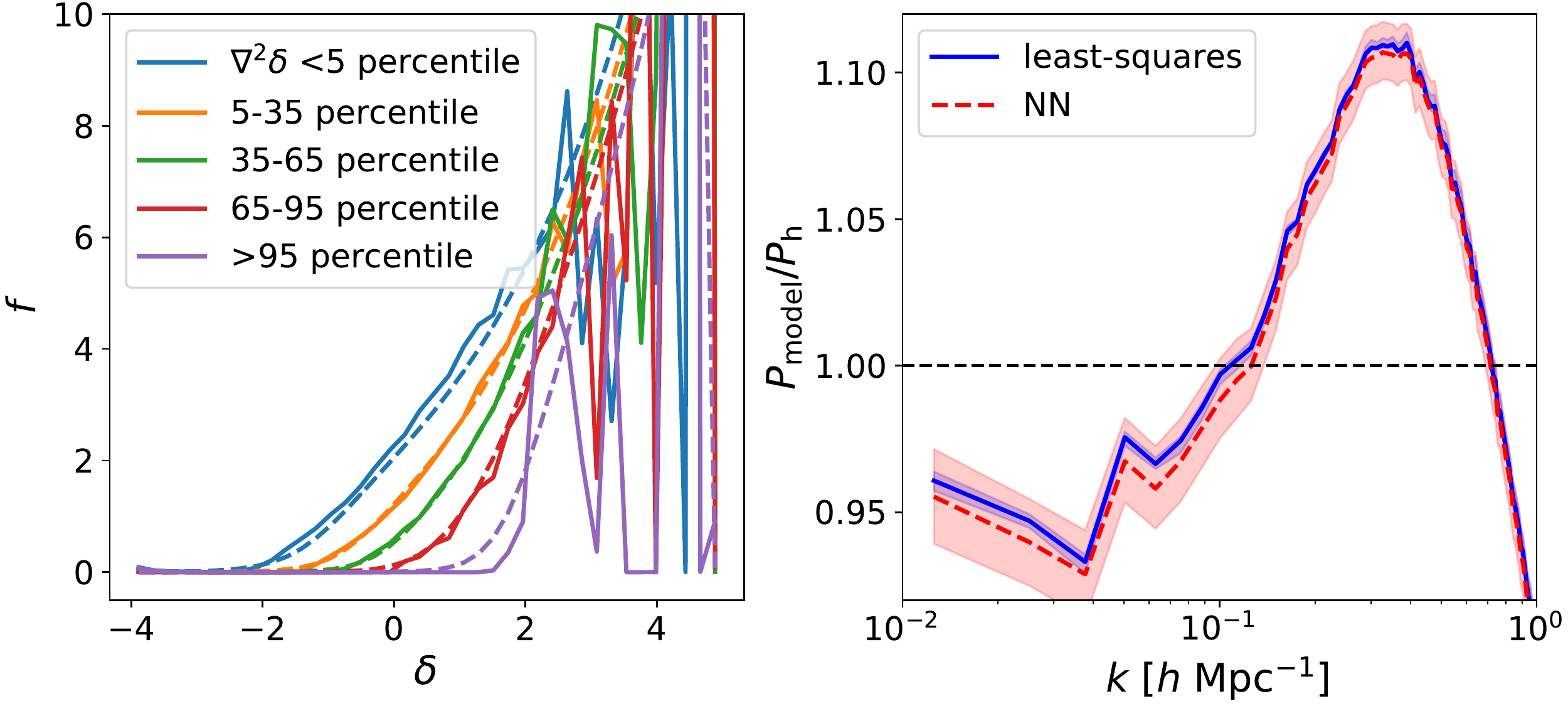}
\caption{Comparison between the least-squares solution obtained from our formalism in Paper I and our NN results on one training simulation.  We use one smoothing scale $R_f=2.83\ h^{-1}$~Mpc and its associated $\delta, \nabla^2\delta$, and trained on one training simulation.  Left panel: $f$ as a function of $\delta$ in different $\nabla^2\delta$ bins, represented by different colors.  Solid and dashed lines show the least-squares and the NN results, respectively.  Right: the ratio of the model power spectrum to the halo power spectrum, evaluated on one validation simulation.  Blue solid and red dashed lines illustrate results from least-squares and NN, respectively.  Shades represent the minimum and maximum ranges obtained from 5 times of training and particle sub-sampling (of the same box), and lines show the results obtained from the average $f$.  The NN recovers the least-squares results well and produces a much smoother $f$.}
\label{fig:compare_lstsq}
\end{figure}

The left panel of Figure~\ref{fig:compare_lstsq} compares the $f$ function obtained from least-squares (solid lines) and from the NN (dashed lines), with the colors indicating five $\nabla^2\delta$ bins.  The NN solution traces the least-squares solution, but is remarkably smoother.  The least-squares solution, on the other hand, shows large noise at high $\delta$ values.  The right panel illustrates the ratio of the model power spectrum to the halo power spectrum $P_{\rm model}/P_{\rm h}$ calculated using one validation simulation.  The solid and dashed lines show $P_{\rm model}$ obtained from the average $f$ function from least-squares and NN, respectively.  Shades represent the minimum and maximum ranges obtained from 5 times of training and random sampling of particles.  Again, the average $f$ of the NNs reproduces the least-squares power spectrum well.

The NN-predicted $f$ yields noticeably larger scatter ($\sim1\%$) in the halo power spectrum at $k<0.1\ h$~Mpc$^{-1}$ when trained five different times, compared to the $\ll1\%$ scatter produced by the least-squares $f$.  Since the least-squares solution is deterministic and noisy but the NN solution is smooth, the NN tries to match the least-squares solution but can never reach the low loss created by least-squares.  A small change in $f$, especially at the high-$\delta$ end, may not affect the real-space loss noticeably but will leave a larger imprint on the predicted bias of the halos.
As we will show below, we find that such 1-2\% fluctuations in power are persistent in all NN results, regardless of the smoothing scales used.  We will examine these scatter in more detail below.

We also note that both the least-squares and the NN solutions underpredict the halo power by $\sim5\%$ at $k\lesssim0.1\ h$~Mpc$^{-1}$.  We showed in Paper I that this deficit can be alleviated by including a high-$k$ cut-off to our loss function at $k=1/R_f$.  We do not perform this cut-off in this work, but rather try to improve the match to the halo field (and indirectly to the power spectrum) by including different smoothing scales.

\subsection{Results with different smoothing scales}

We now compare the NN results with different combinations of smoothing scales, since density fluctuations at different $R_f$'s can contribute to halo collapse with different weights.
The least-square method becomes computationally unfeasible for more than 2 input features since the number of bins grows exponentially with the dimension of the feature spaces, but the NN is able to incorporate a much higher dimensional input space.
For each $R_f$, we include its associated $\delta, \nabla^2\delta, \mathcal{G}_2$ into the input features.  As mentioned in Section~\ref{sec:methods}, we orthogonalize the features and reduce the $3n_{R_f}$-dimensional feature space to a $(2n_{R_f}+1)$-dimensional one based on the eigenvalues, where $n_{R_f}$ is the number of smoothing scales.  For each sequence of $R_f$'s, we make it a geometric series with a common ratio equal to some power of $\sqrt{2}$.  Table~\ref{tab:Rf} summarizes the $R_f$'s that we use and the names referring to them later in the plots.
We use the ratio of the model power spectrum to the halo power spectrum $P_{\rm model}/P_{\rm h}$ as a metric, though we note that it is not included in our loss.

\begin{table}
\centering
\begin{tabular}{c|c}
Name & $R_f$ ($h^{-1}$ Mpc) \\
\hline
1 $R_f$ & $2.83$ \\
3 $R_f$ & $1.41,2.83,5.66$ \\
3 $R_f$ + $11.3$ & $1.41,2.83,5.66,11.3$ \\
3 $R_f$ coarse & $1,2.83,8$ \\
3 $R_f$ fine & $2,2.83,4$ \\
5 $R_f$ fine & $1.41,2,2.83,4,5.66$
\end{tabular}
\caption{The sequences of smoothing scales used for training and their short names.}
\label{tab:Rf}
\end{table}

\begin{figure}
\centering
\includegraphics[width=\linewidth]{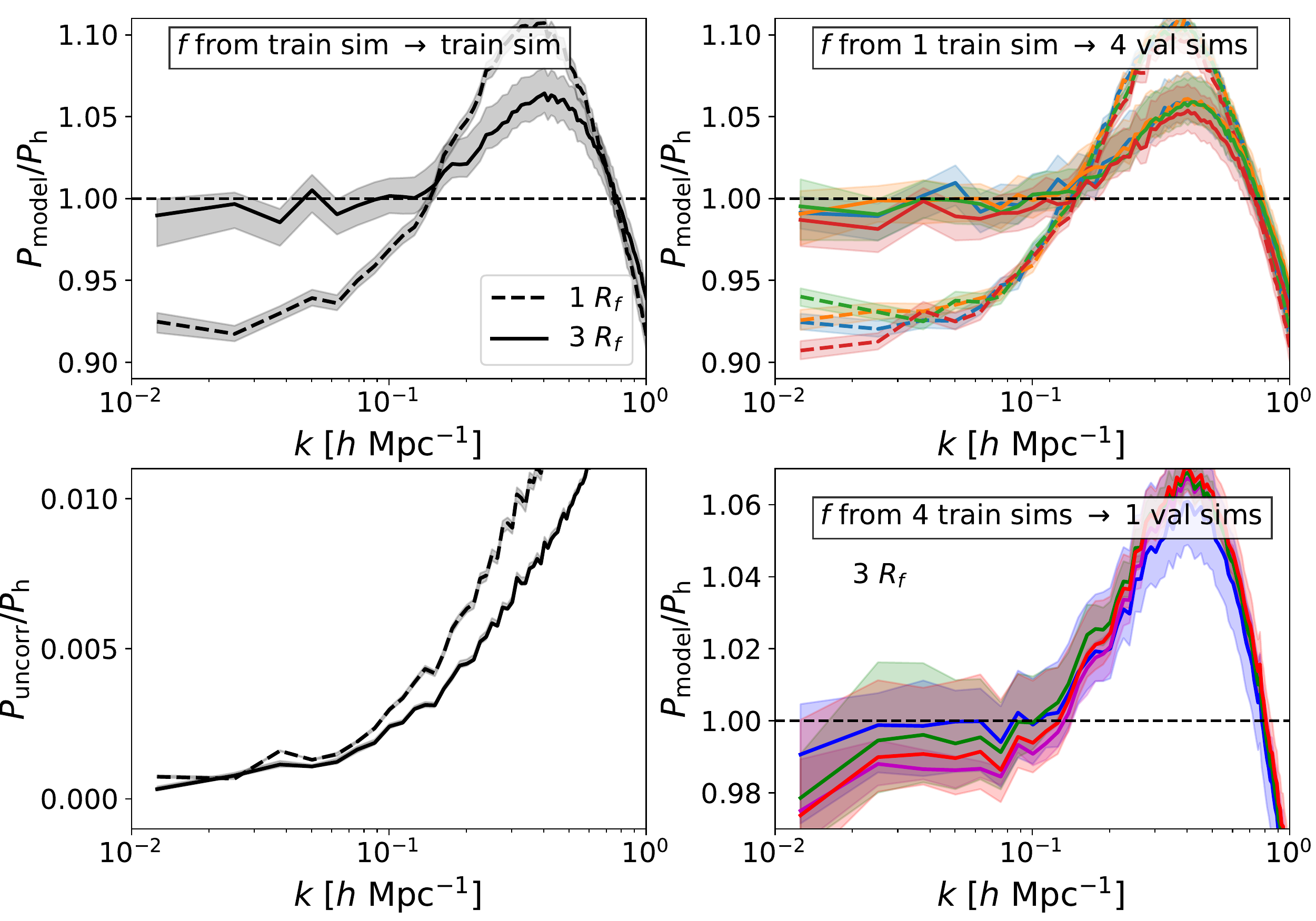}
\caption{Comparison between the training results obtained using one smoothing scale (dashed lines) to using 3 smoothing scales (solid lines), and impact of cosmic variance on the model power spectrum.
Shades represent the minimum and maximum ranges of the power spectra obtained from 5 times of training, and lines show the results obtained from the average $f$.
Left column: results are obtained by applying $f$ from the training simulation to the training simulation.  Top left and bottom left panels show the ratio of the model power spectra to the halo power spectrum, and the ratio of the power spectra of the uncorrelated residual to the halo power spectrum respectively.
Top right panel: $P_{\rm model}/P_{\rm h}$, calculated by applying $f$ from the training simulation to 4 validation simulations (different colors).
Bottom right panel: $P_{\rm model}/P_{\rm h}$, calculated by applying $f$ from 4 training simulations (different colors) to 1 validation simulation.
Adding the two additional smoothing scales significantly improves the match of the halo power spectrum, although cosmic variance leads to 1-2\% scatter in the resulting halo power spectrum.
}
\label{fig:3Rfs_power}
\end{figure}

We first examine the effects of using multiple $R_f$'s instead of one.  Figure~\ref{fig:3Rfs_power} contrasts the training results using $R_f=1.41,2.83,5.66\ h^{-1}$~Mpc (dashed lines) with those of $R_f=2.83\ h^{-1}$~Mpc (solid lines).
Left column shows the result of applying $f$ from the training simulation to the training simulation.  Top left and bottom left panels present $P_{\rm model}/P_{\rm h}$, and the ratio of the power spectra of the uncorrelated residual $P_{\rm uncorr}$ to the halo power spectrum respectively.  Here
\begin{equation}
P_{\rm uncorr} = P_{\rm model} - \frac{P^2_{\rm h,model}}{P_{\rm h}},
\end{equation}
where $P_{\rm h,model}$ is the cross spectrum between the halo field measured and modeled.
This uncorrelated residual characterizes the part in the model halo field that is uncorrelated with the true halo field \cite{wu22, modi17}.

Using 3 smoothing scales dramatically improves the match of $P_{\rm model}$ to $P_{\rm h}$, bringing the ratio to 1 within $\sim2\%$ at $k<0.1\ h$~Mpc$^{-1}$.  With only 1 smoothing scale, we find $P_{\rm model}/P_{\rm h}\lesssim0.95$ at $k<0.1\ h$~Mpc$^{-1}$.  We emphasize that we fit the real-space halo field without filtering the residuals at $k>1/R_f$ in Fourier space as we did in Paper I.
Top right panel shows $P_{\rm model}/P_{\rm h}$, calculated by applying $f$ from the training simulation to 4 validation simulations (different colors).
Results on the validation simulations also trace those of the training simulations, although cosmic variance can lead to up to 5\% scatter in the low-$k$ power of these small boxes.

There are about $1.5\%$ variations in the model power spectrum of individual simulations when using 3 $R_f$, while the one smoothing scale case yields $0.5\%$ scatter in power.  Since $P_{\rm uncorr}/P_{\rm h}<0.002$ at $k<0.1\ h$~Mpc$^{-1}$, all of the variations in $P_{\rm model}$ come from the residuals in the model halo field that is correlated with the true halo field.
We find that whether the model underpredicts or overpredicts the power is uncorrelated with the value of the real-space squared loss.  As mentioned earlier, a small change in $f$, especially at high $\delta$ values, may not affect the real-space loss much since the number of high $\delta^{\rm true}_h$ cells are small.  However, changes in $f$ can leave a more evident imprint on the power spectrum indicating overestimating or underestimating the halo bias.  This issue seems more prominent when the dimension of the input feature space is higher.  This could be alleviated by using a $k$-space loss with some emphasis on the low-$k$ end or by explicitly including in the power spectrum in our fitting.

The fact that our small $500\ h^{-1}$~Mpc box size contains very few low-$k$ modes can also contribute to large low-$k$ scatter.
The bottom right panel shows $P_{\rm model}/P_{\rm h}$ calculated by applying $f$ from 4 training simulations (different colors) to 1 validation simulation.  In addition to the 1-2\% scatter around each power spectrum using the average $f$ from a single training simulation, there is also 1-2\% scatter in the power spectra from the 4 average $f$'s from 4 different training simulations.
This indicates that the fluctuations in the halo power spectrum may be sourced by changes in $f$ owing to cosmic/sample variance.  We thus expect a better match to the power spectrum by training with multiple small boxes at the same time or with a larger box size.
Given the exploratory nature of this first paper, we leave it for future work to make these improvements.

\begin{figure}
\centering
\includegraphics[width=\linewidth]{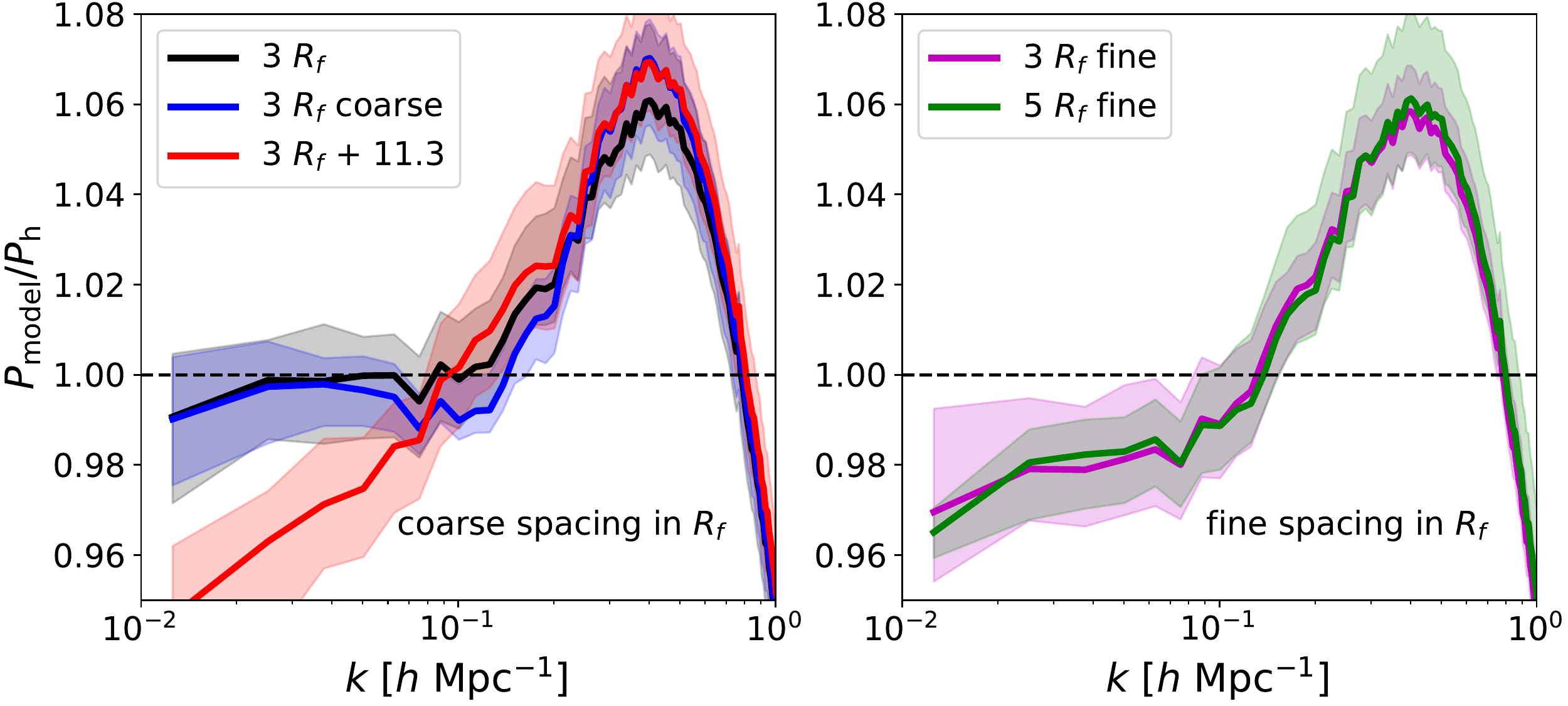}
\caption{Comparison of training results using different combinations of smoothing scales, measured using the ratio of the model power spectra to the halo power spectra, evaluated for one validation simulation.  The smoothing scales are separated by factors of $2$, $2^{3/2}$ (coarse) in the left panels, and $\sqrt{2}$ (fine) in the right, respectively.  Shades represent the minimum and maximum ranges obtained from 5 times of training, and lines show the results obtained from the average $f$.  The black line in the left panel illustrates the $R_f=1.41,2.83,5.66\ h^{-1}$~Mpc case, and the red line shows the result of adding $R_f=11.3\ h^{-1}$~Mpc.  The blue line represents the case of $R_f=1,2.83,8\ h^{-1}$~Mpc.  The magenta and green lines in the right panel show $R_f=2,2.83,4\ h^{-1}$~Mpc and $R_f=1.41,2,2.83,4,5.66\ h^{-1}$~Mpc, respectively.  Coarse-spaced smoothing scales in general result in much better fit to the halo power spectrum, while fine-spaced ones lead to 2-3\% underprediction of the low-$k$ power owing to the highly correlated features making the NN overfits.  Adding in $11.3\ h^{-1}$~Mpc without filtering the residuals in $k$-space gives significantly underestimated low-$k$ power.}
\label{fig:diffRfs_power}
\end{figure}

We now test the impact of using other combinations of smoothing scales, by either adding $R_f$'s or using another splitting of the $R_f$'s.  Here we only evaluate $P_{\rm model}/P_{\rm h}$ by applying $f$ from one training simulation to one validation simulation, since the training and validation power spectra show similar trends.  The left and right panels of Figure~\ref{fig:diffRfs_power} show $P_{\rm model}/P_{\rm h}$ in the case of using coarser (common ratio of 2 or $2^{3/2}$ in $R_f$) and finer (common ratio of $2^{1/2}$) $R_f$'s respectively.  Shades represent the minimum and maximum ranges obtained from 5 times of training, and lines show the results obtained from the average $f$.  The black line in the left panel illustrates the $R_f=1.41,2.83,5.66\ h^{-1}$~Mpc case, and the red line shows the result of adding $R_f=11.3\ h^{-1}$~Mpc to it.  The blue line represents the case of $R_f=1,2.83,8\ h^{-1}$~Mpc.  The magenta and green lines in the right panel show $R_f=2,2.83,4\ h^{-1}$~Mpc and $R_f=1.41,2,2.83,4,5.66\ h^{-1}$~Mpc, respectively.

As mentioned above, all of these cases of training in a very high dimensional feature space lead to about 1-2\% scatter in the model power spectrum.  Using 3 smoothing scales with coarse spacing results in the best match of the power spectrum, and $P_{\rm model}$ agrees with $P_{\rm h}$ to 1\% level.  Adding a large $R_f=11.3\ h^{-1}$~Mpc drastically lowers the low-$k$ power by up to 5\%, which is not reflected in any changes of the loss.  In Paper I, we found that using large smoothing scales with small halo grid cells can lead to a $\sim10\%$ underestimation of the power spectrum, if we do not filter out the high-$k$ residuals.  It is thus likely that the NN outweigh the contribution of the $R_f=11.3\ h^{-1}$~Mpc features which then lowers the low-$k$ power.  Since it is non-trivial to perform the training in Fourier space and determine a proper high-$k$ cut-off in the case of multiple smoothing scales, we will defer for a future work to implement these improvements and just warn the reader about fitting with too high $R_f$'s.

Compared to the coarse-spaced $R_f$'s, which match well the power spectrum, using fine-spaced $R_f$'s yields 2\% underestimation of the power at $k<0.1\ h$~Mpc$^{-1}$.  We find that the real-space squared loss is indistinguishable between the two.  Since the features are more correlated when the $R_f$'s are closer, the orthogonalization and normalization of the original feature space greatly stretches the directions in the feature space that have tiny eigenvalues.  This likely leads to the NN overfitting, in the presence of redundant features.  We will show below, however, that re-training with the first 2 principal components of $\nabla f$ can effectively reduce this redundancy and provide a better match of $P_{\rm model}$ to $P_{\rm h}$, as long as the required $R_f$'s are present.

In summary, we find that the NN-predicted $f$ is much smoother than the least-squares (binned) $f$, and that using features associated with coarse-spaced $R_f$'s leads to the best recovery of the halo power spectrum.  Below we will examine the structure of $f$ in the input feature space in more detail.

\section{Reducing the dimensions of the input feature space}
\label{sec:results_PC}

Our results above showed that expanding the number of coarse-spaced smoothing scales gives a better fit to the halo field.  However, the NN likely overfits in the presence of highly correlated features owing to fine-spaced $R_f$'s.  We now examine whether $f$ can be expressed in terms of a smaller number of input features that constitute a subset of the original input feature space.
In other words, we would like to find if a lower-dimensional subspace of the initial features, beyond the projection already employed, might be used.

As an example, if $f$ depended only upon a linear combination of the input features, we would expect $\nabla f$ to show a unique preferred direction in the feature space, even if $f$ itself was a very nonlinear function.
We therefore explore the structure of $f$ by computing the principal components (PCs) of $\nabla f$.  The PCs are given by the eigenvectors of the covariance matrix of $\nabla f$.  To calculate the covariance matrix, we randomly sample $N_{\rm sample} = 10^6$ particles, perturb around their orthogonalized and normalized features $\theta_{\rm input}$, and calculate their $\nabla f$.  The resulting gradient of $f$ forms a matrix of size $(2n_{R_f}+1)\times N_{\rm sample}$, where $n_{R_f}$ is the number of smoothing scales.  We multiply this $\nabla f$ matrix by its transpose and obtain the eigenvector matrix $U$ of the resulting covariance matrix, where the columns of $U$ are the eigenvectors.  The transformation from the original $\delta, \nabla^2\delta, \mathcal{G}_2$ features to the PCs of $\nabla f$ is thus
\begin{equation}
\theta_{\rm PC} = U^T \theta_{\rm input} = \underbrace{U^T P \Lambda^{-1/2} V^T}_M \theta_{\rm orig},
\label{eq:PC_eq}
\end{equation}
where the second equality follows from equation~\ref{eq:feature_transformation_final}.
Given a set of $R_f$'s, we compute the average $f$ from the five trained NNs and the corresponding PCs of $\nabla f$.  We then examine the structure of $f$ as a function of $\theta_{\rm PC}$.

\begin{figure}
\centering
\includegraphics[width=\linewidth]{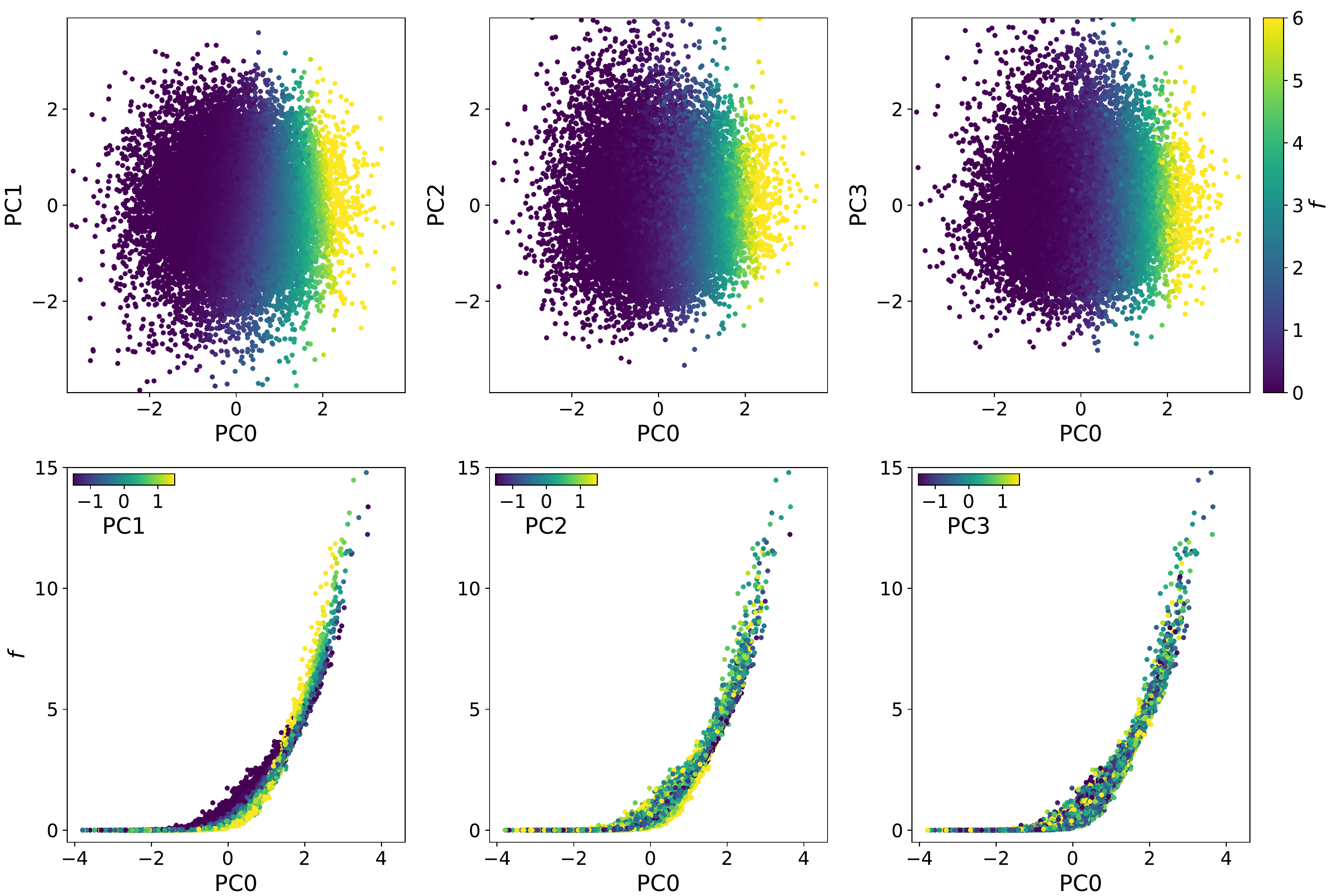}
\caption{Visualization of the $f$ function, obtained by training with $R_f=1.41,2.83,5.66\ h^{-1}$~Mpc in the basis expanded by the principal components (PC) of $\nabla f$.  Top panels: $f$ in the 2D planes of the zeroth, first, second, and third PCs, which have the largest eigenvalues.  Bottom panels: $f$ as a function of the zeroth PC, with the color-bars showing values of the first, second, and third PCs.  Most of the structure of $f$ is contained in the zeroth PC which has the largest eigenvalue, and the dependence of $f$ beyond the first PC becomes rather weak.}
\label{fig:3Rfs_f}
\end{figure}

We use the 3 $R_f$ case as an example, since we find that the behavior of $f$ in the PC space is very similar between the sets of $R_f$'s that we have tested (see Table.~\ref{tab:Rf}).
Figure~\ref{fig:3Rfs_f} shows the visualization of the $f$ function in the PC space.  The top panels illustrate $f$ in the 2D planes of the zeroth, first, second, and third PCs, which have the largest eigenvalues.  Bottom panels presents $f$ as a function of the zeroth PC, with the color-bars showing values of the first, second, and third PCs.

We find that most of the variation in $f$ is carried in the zeroth PC, with the largest eigenvalue.  
Very approximately $f$ is 0 at ${\rm PC0}\lesssim-1$, and increases monotonically with PC0 at larger values.  This behavior of $f$ is very similar to the dependence of $f$ on $\delta$ seen in Paper I, thus making PC0 analogous to the overdensity.  Intuitively, halos form in regions with high initial overdensity.  The higher the halo mass, the larger the required $\delta$.  Since we fit the mass-weighted halo field with a mass threshold, this leads to a monotonically increasing $f$ with $\delta$, or PC0.

\begin{figure}
\centering
\includegraphics[width=\linewidth]{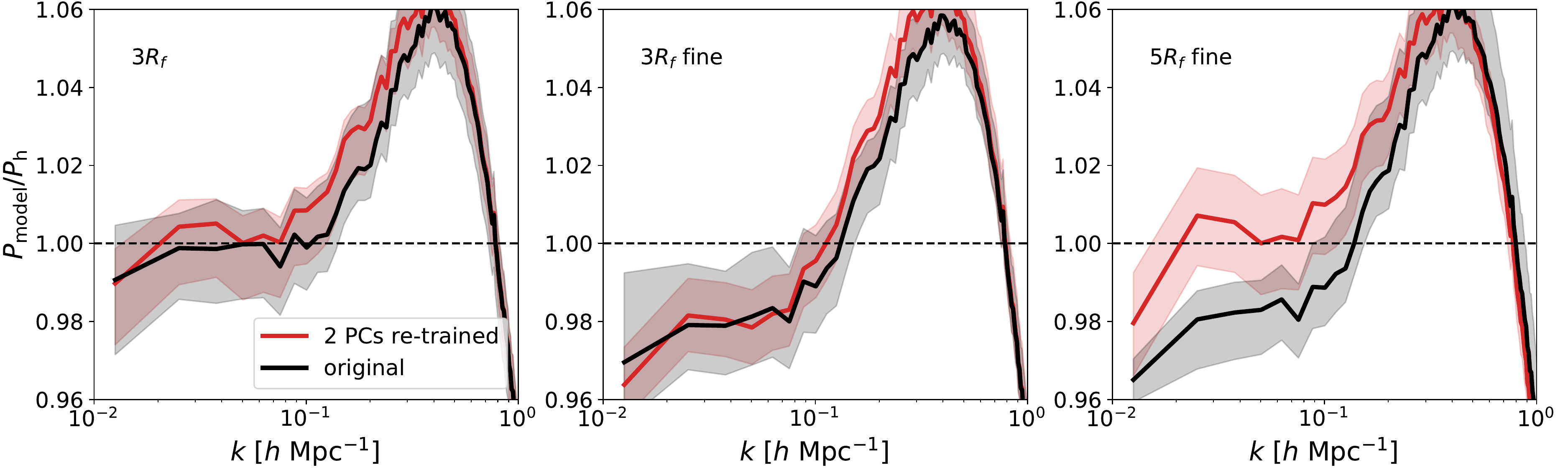}
\caption{Results of re-training the NN after projecting the orthogonalized features onto the directions of the first two PCs of $\nabla f$.  Black and red represent the ratio of the model power spectra to the halo power spectra from the original training results and the re-trained results, respectively.  Shades represent the minimum and maximum ranges obtained from 5 times of training, and lines show the results obtained from the average $f$.  From left to right we show cases of different combinations of $R_f$'s.  Re-training with only 2 PCs either recovers the original training result or outperform it owing to the reduction of the redundancy in the input feature space.}
\label{fig:diffRfs_2PCs_retrain}
\end{figure}

Beyond PC0, additional trend of variation in $f$ can be well explained by PC1, and the dependence of $f$ on the rest of the PCs is much weaker.  We find that regardless of the smoothing scales used for training, the eigenvalue of the zeroth PC is a factor of 3 and 4 larger than those of the first and the second PCs respectively, but the variation of $f$ with PC2 at a fixed PC0 is much weaker than that with PC1.  The rest of the PCs have rather flat amplitudes of the eigenvalues.
We thus examine whether $f$ can be recovered by only two PCs by re-training the NN with the first two rows of $\theta_{\rm PC}$ that correspond to PC0 and PC1 of $\nabla f$.

Figure~\ref{fig:diffRfs_2PCs_retrain} show $P_{\rm model}/P_{\rm h}$ after the re-training with PC0 and PC1.  Black and red colors represent the original training results and the re-trained results, respectively.  Shades represent the minimum and maximum ranges obtained from 5 times of training, and lines show the results obtained from the average $f$.  From left to right we show cases of different combinations of $R_f$'s: $R_f=1.41,2.83,5.66\ h^{-1}$~Mpc, $R_f=2,2.83,4\ h^{-1}$~Mpc, $R_f=1.41,2,2.83,4,5.66\ h^{-1}$~Mpc.  Although not shown, we find that re-training with the first two PCs also recovers the original training and validation losses.

For $R_f=1.41,2.83,5.66\ h^{-1}$~Mpc and $R_f=2,2.83,4\ h^{-1}$~Mpc, the re-training produces $P_{\rm model}/P_{\rm h}\approx1$ and $0.98$ at $k<0.1\ h$~Mpc$^{-1}$ respectively, the same as the original training results.  This demonstrates that the first two PCs are adequate to characterize $f$.  For $R_f=1.41,2,2.83,4,5.66\ h^{-1}$~Mpc, the re-trained results bring $P_{\rm model}/P_{\rm h}$ close to 1, while this ratio is $0.98$ in the original training.  This indicates that by reducing the size of the input feature space with only the first two PCs of $\nabla f$, the NN no longer overfits owing to the redundancy in the highly correlated features.  For the case of $R_f=1.41,2.83,5.66,11.3\ h^{-1}$~Mpc, although not shown here, we find that the re-training raises the low-$k$ power by 3\% and brings $P_{\rm model}/P_{\rm h}$ to $0.98$ at low-$k$.  The re-training thus improves the match to the power spectrum, in the absence of filtering the residuals.

\begin{figure}
\centering
\includegraphics[width=\linewidth]{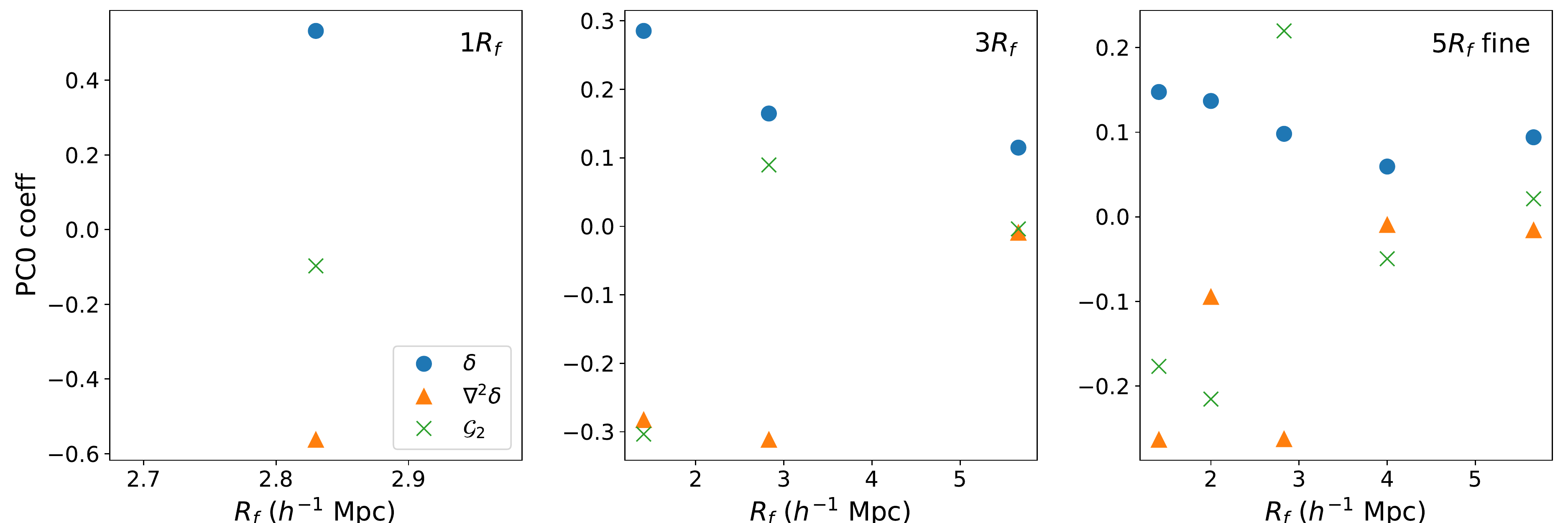}
\caption{Coefficients of the zeroth principal component of $\nabla f$ at corresponding values of $R_f$.  From left to right, the NN is trained with $R_f=2.83\ h^{-1}$~Mpc, $R_f=1.41, 2.83, 5.66\ h^{-1}$~Mpc, $R_f=1.41, 2, 2.83, 4, 5.66\ h^{-1}$~Mpc, respectively.
Blue dots, orange triangles, and green crosses represent $\delta, \nabla^2\delta, \mathcal{G}_2$ respectively.
There is hardly any specific structure in the coefficients, so it is not desirable to interpret the numbers with any physical meanings.}
\label{fig:diffRfs_PC0_coeff}
\end{figure}

We now examine whether the elements of the transformation matrix $M$, which we term the ``coefficients of the PCs'', have any physical implications.  Intuitively, these numbers might be interpreted as the bias coefficients.
Figure~\ref{fig:diffRfs_PC0_coeff} shows coefficients of PC0 of $\nabla f$ at corresponding $R_f$'s.  From left to right, the NN is trained with $R_f=2.83\ h^{-1}$~Mpc, $R_f=1.41, 2.83, 5.66\ h^{-1}$~Mpc, $R_f=1.41, 2, 2.83, 4, 5.66\ h^{-1}$~Mpc, respectively.  Blue dots, orange triangles, and green crosses represent the coefficients of $\delta, \nabla^2\delta, \mathcal{G}_2$ respectively.

The one smoothing scale case produces a close-to-zero coefficient for $\mathcal{G}_2$.  This confirms our previous findings in Paper I that $\mathcal{G}_2$ is not as important as $\nabla^2\delta$ in recovering the halo field with a cut at $M>3\times10^{11}\ M_\odot$.  Previous works also find that either the tidal bias is only important for halos with $M\gtrsim10^{13}\ h^{-1}\ M_\odot$ \cite{abidi18}, or there is a small negative shear bias regardless of halo mass \cite{bel15, lazeyras18}.
When including more smoothing scales, the coefficients of $\mathcal{G}_2$ jump up and down around zero and can have larger amplitudes than the coefficients of $\delta$ and $\nabla^2\delta$.  This is partially caused by the orthogonalization and normalization of the original features introducing small values in $\Lambda$, which when inverted and multiplied with $U$ enlarges the corresponding $U$ elements (equation~\ref{eq:PC_eq}).  In fact, the elements in $U$ that are associated with the orthogonalized $\mathcal{G}_2$'s have a factor of 2-3 lower amplitude than those associated with the orthogonalized $\delta$ and $\nabla^2\delta$.  Moreover, since the $\mathcal{G}_2$'s are correlated, the jumping above and below zero of the $\mathcal{G}_2$ coefficients reflects the cancellation of the effects of the $\mathcal{G}_2$ terms.
Our results with multiple smoothing scales thus are still consistent with $\mathcal{G}_2$ playing a minor role in determining the halo field with $M>3\times10^{11}\ h^{-1}\ M_\odot$.

Comparing the $3 R_f$ case and $5 R_f$ one, the coefficients of $\delta$ and $\nabla^2\delta$ are similar at the $R_f$ values of common.  The addition of two intermediate smoothing scales brings in values that continues the spectrum of the coefficients of $\delta$, but the spectrum of the coefficients of $\nabla^2\delta$ seems less regular.  We find similar trends when examining the case of $R_f=1.41, 2.83, 5.66, 11.3\ h^{-1}$~Mpc and $R_f=1, 2.83, 8\ h^{-1}$~Mpc.  It is thus unclear whether these coefficients can be interpreted with any physical meaning.  It might be possible to derive the spectrum of coefficients using the peak-patch formalism or the extended Press-Schechter theory, but such an exploration is beyond the scope of our paper.  We leave it for a future work to examine this in detail.

In summary, we find that $f$ can be well described by two directions in the original features space, and that training with these two principal components prevents the NN from overfitting.  While previous works using the bias expansion approach to model the halo field only use one smoothing scale \cite[e.g.][]{modi17, lazeyras16, lazeyras18, lazeyras19, abidi18, zennaro21}, our results indicate that a linear combination of the features of multiple smoothing scales may lead to a better fit of the halo field.

\section{Modeling a thin mass range}
\label{sec:thin_mass_range}

Having set up all the necessary tools for fitting $f$ and examining its structure, we now briefly explore fitting the mass-weighted halo field with a thin mass range instead of a mass threshold.  Intuitively, the larger the halo mass, the more it requires the involvement of higher $\delta$ regions.  To obtain halos within a thin mass range, regions with too high $\delta$ thus cannot participate. As a consequence, instead of monotonically increasing with $\delta$, as in the case of using mass thresholds, $f$ will rise with $\delta$ but eventually drop to zero \cite{matsubara08}.  We also expect the peak of $f$ to shift towards higher $\delta$ for higher-mass bins.

To test these intuitions, we train NNs on the mass-weighted halo fields with $3\times10^{11}\ M_\odot < M < 10^{12}\ M_\odot$ and $2\times10^{12}\ M_\odot < M < 10^{13}\ M_\odot$.  The mass-weighted halo masses are $5.2\times10^{11}\ M_\odot$ and $6.2\times10^{12}\ M_\odot$, corresponding to Gaussian filters with $R_f=0.8\ h^{-1}$~Mpc and $1.9\ h^{-1}$~Mpc, respectively.  Since our initial density field is created on a grid with cell size of $0.87\ h^{-1}$~Mpc, we only produce smoothed density fields with $R_f\ge1\ h^{-1}$~Mpc.  However, we expect the case of fitting a thin mass bin to require more strongly the involvement of multiple smoothing scales.  Intuitively, the overdensity smoothed at the halo scale $R_h$ should be large in order to form halos within a thin mass range, but the overdensity smoothed at $R_f \gg R_h$ needs to be small enough to not collapse to bigger objects.  The NN therefore needs intake of features at a wide range of $R_f$'s.
We thus train the NN with the $\delta,\nabla^2\delta,\mathcal{G}_2$'s at $R_f=1,2,4\ h^{-1}$~Mpc, which correspond to typical masses $M_h=1.4\times10^{12},1.1\times10^{13},8.7\times10^{13}\ M_\odot$.  As above, we perform five times of training on one training simulation with different initializations and obtain the average $f$.

\begin{figure}
\centering
\includegraphics[width=\linewidth]{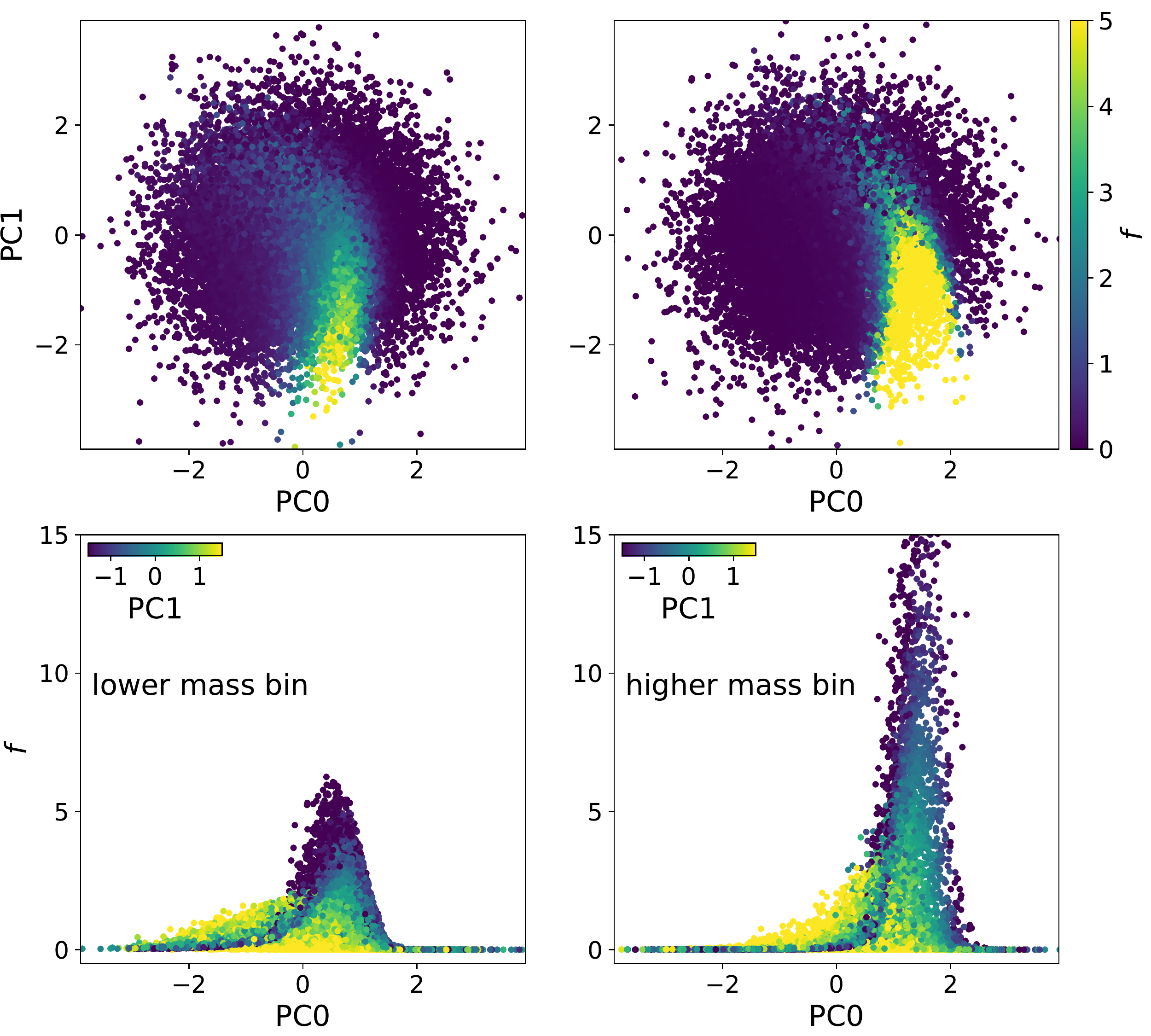}
\caption{Visualization of the $f$ function on two thin mass ranges, defined by $3\times10^{11}\ M_\odot < M < 10^{12}\ M_\odot$ (left) and $2\times10^{12}\ M_\odot < M < 10^{13}\ M_\odot$ (right), obtained by training with $R_f=1,2,4\ h^{-1}$~Mpc.  Top panels: $f$ in the 2D planes of the zeroth and first PCs.  Bottom panels: $f$ as a function of the zeroth PC, with the color-bars showing values of the first PC.  The structure of $f$ in the case of a thin mass range is substantially different from training with a mass threshold.  The peak of $f$ shifts to higher values of PC0 as the halos become more massive, as PC0 is analogous to the overdensity.}
\label{fig:3Rfs_f_massbin}
\end{figure}

We calculate the PCs of $\nabla f$ and examine $f$ as a function of the PCs.  Figure~\ref{fig:3Rfs_f_massbin} shows $f$ of the lower (left) and higher (right) mass range halos.  The top panels illustrate $f$ in the 2D planes of the zeroth and first PCs.  The bottom panels present $f$ as a function of the zeroth PC, with the color-bars showing values of the first PC.

As expected, $f$ in the case of a thin mass range rises and drops with increasing values of PC0, substantially different from the case of a mass threshold where $f$ increases monotonically with PC0.  The peak of $f$ occurs at ${\rm PC0}\sim 0.5$ for the lower-mass halos, but at ${\rm PC0}\sim 1.5$ for the higher-mass ones.  As PC0 is analogous to the overdensity, this matches our intuition that higher overdensities contribute more to the formation of more massive halos.  $f$ also exhibits a tilt in the PC0-PC1 plane, compared to its more regular structure in the case of mass threshold (Figure~\ref{fig:3Rfs_f}).  These complex structures of $f$ demonstrate the necessity of using NNs to explore halo formation in a large parameter space.  We will examine the training on thin-mass-range halos in detail in a future work, including the performance of NNs on matching the halo power spectrum.

\section{Conclusions}
\label{sec:conclusions}

In this work we train neural nets to obtain a non-parametric Lagrangian model of galaxy clustering bias, expanding our previous formalism that uses least-square fitting on a binned input feature space.  We measure the halo-to-mass ratios $f$ for mass-weighted halos in N-body simulations, assuming $f$ depends on the linear overdensity $\delta$, the tidal operator $\mathcal{G}_2$, and a non-local term $\nabla^2\delta$, all of which are smoothed over multiple smoothing scales.
We find that for mass-weighted halos above a mass threshold, using 3 coarsely separated smoothing scales gives a much better recovery of the halo power spectrum than 1 smoothing scale that corresponds to the halo mass.  Adding more smoothing scales or using fine-spaced ones leads to overfitting and thus a 2-5\% underestimation of the low-$k$ power.  By calculating the principal components (PC) of $\nabla f$, we find that $f$ can be well described by the first two PCs of $\nabla f$ and that $f$ is a monotonically increasing function of the PC with the largest eigenvalue (PC0).  Re-training the NN with these two PCs either recovers the original training results or outperforms it by better matching the halo power spectrum, indicating that they prevent the NN from overfitting.  The coefficients of these linear combinations may be interpreted as the bias in the case of multiple smoothing scales, but a detailed examination is beyond the scope of our paper.

We briefly explored fitting the mass-weighted halo field over a thin mass range instead of a mass threshold, and find that $f$ rises and drops with PC0 instead of being monotonically increasing.  This matches our physical intuition that $f$ should only be non-zero at a certain range of overdensities in the case of a thin mass range.  The complex structure of $f$ in the PC space demonstrates the usefulness of using a NN to examine structure formation.

As we find that the real-space squared loss is not correlated with the recovery of the power spectrum, one way to improve our formalism is to express the loss function in Fourier space and filtering out high-$k$ residuals.  Producing and fitting the halo fields using Cloud-in-Cell interpolation with varying kernel size instead of nearest neighbor interpolation may have a similar effect as filtering the high-$k$ residuals and thus can also help alleviate the difficulty of matching the low-$k$ power.
We will also extend our code to perform the fitting in $2 h^{-1}$~Gpc large box simulations, which may largely reduce the 1-2\% scatter in the model power spectrum.

Despite all these future improvements, we have demonstrated the ability of using a NN to predict complex-structured $f$ in a high-dimensional input feature space, and that multiple smoothing scales are needed to fully capture the halo field.
This is especially true in the thin mass range case.  We will explore in a future work the use of number-weighted halos or a halo occupation distribution model \cite{yuan18, hadzhiyska20, hadzhiyska21}, which should also require complicated structures of $f$.
Another interesting direction of future work is to examine $f$ of halos at higher redshifts, as these are more biased than the $z=0.5$ halos we study here.
These insights might transform the way we analyze observational data from galaxy surveys, especially those targeting higher redshifts and larger halo masses.

\acknowledgments
JBM is supported by a Clay fellowship at the Smithsonian Astrophysical Observatory. 
DJE is supported by U.S. Department of Energy grant, now DE-SC0007881, by the National Science Foundation under Cooperative Agreement PHY-2019786 (the NSF AI Institute for Artificial Intelligence and Fundamental Interactions, http://iaifi.org/), and as a Simons Foundation Investigator.

\appendix

\paragraph{Note added.} 

\bibliography{References}

\providecommand{\href}[2]{#2}\begingroup\raggedright\begin{thebibliography}{10}

\bibitem{desjacques18}
V.~{Desjacques}, D.~{Jeong} and F.~{Schmidt}, \emph{{Large-scale galaxy bias}},
  \href{https://doi.org/10.1016/j.physrep.2017.12.002}{\emph{\physrep}
  {\bfseries 733} (2018) 1} [\href{https://arxiv.org/abs/1611.09787}{{\ttfamily
  1611.09787}}].

\bibitem{fry93}
J.~N. {Fry} and E.~{Gaztanaga}, \emph{{Biasing and Hierarchical Statistics in
  Large-Scale Structure}}, \href{https://doi.org/10.1086/173015}{\emph{\apj}
  {\bfseries 413} (1993) 447}
  [\href{https://arxiv.org/abs/astro-ph/9302009}{{\ttfamily
  astro-ph/9302009}}].

\bibitem{matsubara08}
T.~{Matsubara}, \emph{{Nonlinear perturbation theory with halo bias and
  redshift-space distortions via the Lagrangian picture}},
  \href{https://doi.org/10.1103/PhysRevD.78.083519}{\emph{\prd} {\bfseries 78}
  (2008) 083519} [\href{https://arxiv.org/abs/0807.1733}{{\ttfamily
  0807.1733}}].

\bibitem{vlah16}
Z.~{Vlah}, E.~{Castorina} and M.~{White}, \emph{{The Gaussian streaming model
  and convolution Lagrangian effective field theory}},
  \href{https://doi.org/10.1088/1475-7516/2016/12/007}{\emph{\jcap} {\bfseries
  2016} (2016) 007} [\href{https://arxiv.org/abs/1609.02908}{{\ttfamily
  1609.02908}}].

\bibitem{chan12}
K.~C. {Chan}, R.~{Scoccimarro} and R.~K. {Sheth}, \emph{{Gravity and
  large-scale nonlocal bias}},
  \href{https://doi.org/10.1103/PhysRevD.85.083509}{\emph{\prd} {\bfseries 85}
  (2012) 083509} [\href{https://arxiv.org/abs/1201.3614}{{\ttfamily
  1201.3614}}].

\bibitem{baldauf12}
T.~{Baldauf}, U.~{Seljak}, V.~{Desjacques} and P.~{McDonald}, \emph{{Evidence
  for quadratic tidal tensor bias from the halo bispectrum}},
  \href{https://doi.org/10.1103/PhysRevD.86.083540}{\emph{\prd} {\bfseries 86}
  (2012) 083540} [\href{https://arxiv.org/abs/1201.4827}{{\ttfamily
  1201.4827}}].

\bibitem{saito14}
S.~{Saito}, T.~{Baldauf}, Z.~{Vlah}, U.~{Seljak}, T.~{Okumura} and
  P.~{McDonald}, \emph{{Understanding higher-order nonlocal halo bias at large
  scales by combining the power spectrum with the bispectrum}},
  \href{https://doi.org/10.1103/PhysRevD.90.123522}{\emph{\prd} {\bfseries 90}
  (2014) 123522} [\href{https://arxiv.org/abs/1405.1447}{{\ttfamily
  1405.1447}}].

\bibitem{abidi18}
M.~M. {Abidi} and T.~{Baldauf}, \emph{{Cubic halo bias in Eulerian and
  Lagrangian space}},
  \href{https://doi.org/10.1088/1475-7516/2018/07/029}{\emph{\jcap} {\bfseries
  2018} (2018) 029} [\href{https://arxiv.org/abs/1802.07622}{{\ttfamily
  1802.07622}}].

\bibitem{fujita20}
T.~{Fujita}, V.~{Mauerhofer}, L.~{Senatore}, Z.~{Vlah} and R.~{Angulo},
  \emph{{Very massive tracers and higher derivative biases}},
  \href{https://doi.org/10.1088/1475-7516/2020/01/009}{\emph{\jcap} {\bfseries
  2020} (2020) 009} [\href{https://arxiv.org/abs/1609.00717}{{\ttfamily
  1609.00717}}].

\bibitem{modi20}
C.~{Modi}, S.-F. {Chen} and M.~{White}, \emph{{Simulations and symmetries}},
  \href{https://doi.org/10.1093/mnras/staa251}{\emph{\mnras} {\bfseries 492}
  (2020) 5754} [\href{https://arxiv.org/abs/1910.07097}{{\ttfamily
  1910.07097}}].

\bibitem{kokron21}
N.~{Kokron}, J.~{DeRose}, S.-F. {Chen}, M.~{White} and R.~H. {Wechsler},
  \emph{{The cosmology dependence of galaxy clustering and lensing from a
  hybrid $N$-body-perturbation theory model}}, {\emph{arXiv e-prints} (2021)
  arXiv:2101.11014} [\href{https://arxiv.org/abs/2101.11014}{{\ttfamily
  2101.11014}}].

\bibitem{schmittfull19}
M.~{Schmittfull}, M.~{Simonovi{\'c}}, V.~{Assassi} and M.~{Zaldarriaga},
  \emph{{Modeling biased tracers at the field level}},
  \href{https://doi.org/10.1103/PhysRevD.100.043514}{\emph{\prd} {\bfseries
  100} (2019) 043514} [\href{https://arxiv.org/abs/1811.10640}{{\ttfamily
  1811.10640}}].

\bibitem{schmittfull20}
M.~{Schmittfull}, M.~{Simonovi{\'c}}, M.~M. {Ivanov}, O.~H.~E. {Philcox} and
  M.~{Zaldarriaga}, \emph{{Modeling Galaxies in Redshift Space at the Field
  Level}}, {\emph{arXiv e-prints} (2020) arXiv:2012.03334}
  [\href{https://arxiv.org/abs/2012.03334}{{\ttfamily 2012.03334}}].

\bibitem{modi19}
C.~{Modi}, M.~{White}, A.~{Slosar} and E.~{Castorina}, \emph{{Reconstructing
  large-scale structure with neutral hydrogen surveys}},
  \href{https://doi.org/10.1088/1475-7516/2019/11/023}{\emph{\jcap} {\bfseries
  2019} (2019) 023} [\href{https://arxiv.org/abs/1907.02330}{{\ttfamily
  1907.02330}}].

\bibitem{barreira21}
A.~{Barreira}, T.~{Lazeyras} and F.~{Schmidt}, \emph{{Galaxy bias from forward
  models: linear and second-order bias of IllustrisTNG galaxies}}, {\emph{arXiv
  e-prints} (2021) arXiv:2105.02876}
  [\href{https://arxiv.org/abs/2105.02876}{{\ttfamily 2105.02876}}].

\bibitem{kaiser84}
N.~{Kaiser}, \emph{{On the spatial correlations of Abell clusters.}},
  \href{https://doi.org/10.1086/184341}{\emph{\apjl} {\bfseries 284} (1984)
  L9}.

\bibitem{desjacques10}
V.~{Desjacques}, M.~{Crocce}, R.~{Scoccimarro} and R.~K. {Sheth},
  \emph{{Modeling scale-dependent bias on the baryonic acoustic scale with the
  statistics of peaks of Gaussian random fields}},
  \href{https://doi.org/10.1103/PhysRevD.82.103529}{\emph{\prd} {\bfseries 82}
  (2010) 103529} [\href{https://arxiv.org/abs/1009.3449}{{\ttfamily
  1009.3449}}].

\bibitem{musso12}
M.~{Musso}, A.~{Paranjape} and R.~K. {Sheth}, \emph{{Scale-dependent halo bias
  in the excursion set approach}},
  \href{https://doi.org/10.1111/j.1365-2966.2012.21903.x}{\emph{\mnras}
  {\bfseries 427} (2012) 3145}
  [\href{https://arxiv.org/abs/1205.3401}{{\ttfamily 1205.3401}}].

\bibitem{baldauf15}
T.~{Baldauf}, V.~{Desjacques} and U.~{Seljak}, \emph{{Velocity bias in the
  distribution of dark matter halos}},
  \href{https://doi.org/10.1103/PhysRevD.92.123507}{\emph{\prd} {\bfseries 92}
  (2015) 123507} [\href{https://arxiv.org/abs/1405.5885}{{\ttfamily
  1405.5885}}].

\bibitem{modi17}
C.~{Modi}, E.~{Castorina} and U.~{Seljak}, \emph{{Halo bias in Lagrangian
  space: estimators and theoretical predictions}},
  \href{https://doi.org/10.1093/mnras/stx2148}{\emph{\mnras} {\bfseries 472}
  (2017) 3959} [\href{https://arxiv.org/abs/1612.01621}{{\ttfamily
  1612.01621}}].

\bibitem{lazeyras16}
T.~{Lazeyras}, C.~{Wagner}, T.~{Baldauf} and F.~{Schmidt}, \emph{{Precision
  measurement of the local bias of dark matter halos}},
  \href{https://doi.org/10.1088/1475-7516/2016/02/018}{\emph{\jcap} {\bfseries
  2016} (2016) 018} [\href{https://arxiv.org/abs/1511.01096}{{\ttfamily
  1511.01096}}].

\bibitem{lazeyras18}
T.~{Lazeyras} and F.~{Schmidt}, \emph{{Beyond LIMD bias: a measurement of the
  complete set of third-order halo bias parameters}},
  \href{https://doi.org/10.1088/1475-7516/2018/09/008}{\emph{\jcap} {\bfseries
  2018} (2018) 008} [\href{https://arxiv.org/abs/1712.07531}{{\ttfamily
  1712.07531}}].

\bibitem{lazeyras19}
T.~{Lazeyras} and F.~{Schmidt}, \emph{{A robust measurement of the first
  higher-derivative bias of dark matter halos}},
  \href{https://doi.org/10.1088/1475-7516/2019/11/041}{\emph{\jcap} {\bfseries
  2019} (2019) 041} [\href{https://arxiv.org/abs/1904.11294}{{\ttfamily
  1904.11294}}].

\bibitem{lazeyras21}
T.~{Lazeyras}, A.~{Barreira} and F.~{Schmidt}, \emph{{Assembly bias in
  quadratic bias parameters of dark matter halos from forward modeling}},
  {\emph{arXiv e-prints} (2021) arXiv:2106.14713}
  [\href{https://arxiv.org/abs/2106.14713}{{\ttfamily 2106.14713}}].

\bibitem{wu22}
X.~{Wu}, J.~B. {Mu{\~n}oz} and D.~{Eisenstein}, \emph{{A fully Lagrangian,
  non-parametric bias model for dark matter halos}},
  \href{https://doi.org/10.1088/1475-7516/2022/02/002}{\emph{\jcap} {\bfseries
  2022} (2022) 002} [\href{https://arxiv.org/abs/2109.13948}{{\ttfamily
  2109.13948}}].

\bibitem{mcdonald09}
P.~{McDonald} and A.~{Roy}, \emph{{Clustering of dark matter tracers:
  generalizing bias for the coming era of precision LSS}},
  \href{https://doi.org/10.1088/1475-7516/2009/08/020}{\emph{\jcap} {\bfseries
  2009} (2009) 020} [\href{https://arxiv.org/abs/0902.0991}{{\ttfamily
  0902.0991}}].

\bibitem{assassi14}
V.~{Assassi}, D.~{Baumann}, D.~{Green} and M.~{Zaldarriaga},
  \emph{{Renormalized halo bias}},
  \href{https://doi.org/10.1088/1475-7516/2014/08/056}{\emph{\jcap} {\bfseries
  2014} (2014) 056} [\href{https://arxiv.org/abs/1402.5916}{{\ttfamily
  1402.5916}}].

\bibitem{bardeen86}
J.~M. {Bardeen}, J.~R. {Bond}, N.~{Kaiser} and A.~S. {Szalay}, \emph{{The
  Statistics of Peaks of Gaussian Random Fields}},
  \href{https://doi.org/10.1086/164143}{\emph{\apj} {\bfseries 304} (1986) 15}.

\bibitem{mo96}
H.~J. {Mo} and S.~D.~M. {White}, \emph{{An analytic model for the spatial
  clustering of dark matter haloes}},
  \href{https://doi.org/10.1093/mnras/282.2.347}{\emph{\mnras} {\bfseries 282}
  (1996) 347} [\href{https://arxiv.org/abs/astro-ph/9512127}{{\ttfamily
  astro-ph/9512127}}].

\bibitem{sheth99}
R.~K. {Sheth} and G.~{Tormen}, \emph{{Large-scale bias and the peak background
  split}},
  \href{https://doi.org/10.1046/j.1365-8711.1999.02692.x}{\emph{\mnras}
  {\bfseries 308} (1999) 119}
  [\href{https://arxiv.org/abs/astro-ph/9901122}{{\ttfamily
  astro-ph/9901122}}].

\bibitem{luciesmith18}
L.~{Lucie-Smith}, H.~V. {Peiris}, A.~{Pontzen} and M.~{Lochner}, \emph{{Machine
  learning cosmological structure formation}},
  \href{https://doi.org/10.1093/mnras/sty1719}{\emph{\mnras} {\bfseries 479}
  (2018) 3405} [\href{https://arxiv.org/abs/1802.04271}{{\ttfamily
  1802.04271}}].

\bibitem{luciesmith19}
L.~{Lucie-Smith}, H.~V. {Peiris} and A.~{Pontzen}, \emph{{An interpretable
  machine-learning framework for dark matter halo formation}},
  \href{https://doi.org/10.1093/mnras/stz2599}{\emph{\mnras} {\bfseries 490}
  (2019) 331} [\href{https://arxiv.org/abs/1906.06339}{{\ttfamily
  1906.06339}}].

\bibitem{luciesmith20}
L.~{Lucie-Smith}, H.~V. {Peiris}, A.~{Pontzen}, B.~{Nord} and
  J.~{Thiyagalingam}, \emph{{Deep learning insights into cosmological structure
  formation}}, {\emph{arXiv e-prints} (2020) arXiv:2011.10577}
  [\href{https://arxiv.org/abs/2011.10577}{{\ttfamily 2011.10577}}].

\bibitem{maksimova21}
N.~A. {Maksimova}, L.~H. {Garrison}, D.~J. {Eisenstein}, B.~{Hadzhiyska},
  S.~{Bose} and T.~P. {Satterthwaite}, \emph{{ABACUSSUMMIT: A Massive Set of
  High-Accuracy, High-Resolution N-Body Simulations}},
  \href{https://doi.org/10.1093/mnras/stab2484}{\emph{\mnras} (2021)
  {\color{black}accepted}}.

\bibitem{garrison18}
L.~H. {Garrison}, D.~J. {Eisenstein}, D.~{Ferrer}, J.~L. {Tinker}, P.~A.
  {Pinto} and D.~H. {Weinberg}, \emph{{The Abacus Cosmos: A Suite of
  Cosmological N-body Simulations}},
  \href{https://doi.org/10.3847/1538-4365/aabfd3}{\emph{\apjs} {\bfseries 236}
  (2018) 43} [\href{https://arxiv.org/abs/1712.05768}{{\ttfamily 1712.05768}}].

\bibitem{garrison19}
L.~H. {Garrison}, D.~J. {Eisenstein} and P.~A. {Pinto}, \emph{{A high-fidelity
  realization of the Euclid code comparison N-body simulation with ABACUS}},
  \href{https://doi.org/10.1093/mnras/stz634}{\emph{\mnras} {\bfseries 485}
  (2019) 3370} [\href{https://arxiv.org/abs/1810.02916}{{\ttfamily
  1810.02916}}].

\bibitem{garrison21}
L.~H. {Garrison}, D.~J. {Eisenstein}, D.~{Ferrer}, N.~A. {Maksimova} and P.~A.
  {Pinto}, \emph{{The ABACUS cosmological N-body code}},
  \href{https://doi.org/10.1093/mnras/stab2482}{\emph{\mnras} (2021)
  {\color{black}accepted}}.

\bibitem{metchnik09}
M.~V.~L. {Metchnik}, \emph{{A fast N-body scheme for computational cosmology}},
  Ph.D. thesis, The University of Arizona, Jan., 2009.

\bibitem{desi1}
{DESI Collaboration}, A.~{Aghamousa}, J.~{Aguilar}, S.~{Ahlen}, S.~{Alam},
  L.~E. {Allen} et~al., \emph{{The DESI Experiment Part I: Science,Targeting,
  and Survey Design}}, {\emph{arXiv e-prints} (2016) arXiv:1611.00036}
  [\href{https://arxiv.org/abs/1611.00036}{{\ttfamily 1611.00036}}].

\bibitem{planck18}
{Planck Collaboration}, N.~{Aghanim}, Y.~{Akrami}, M.~{Ashdown}, J.~{Aumont},
  C.~{Baccigalupi} et~al., \emph{{Planck 2018 results. VI. Cosmological
  parameters}}, \href{https://doi.org/10.1051/0004-6361/201833910}{\emph{\aap}
  {\bfseries 641} (2020) A6}
  [\href{https://arxiv.org/abs/1807.06209}{{\ttfamily 1807.06209}}].

\bibitem{hadzhiyska22}
B.~{Hadzhiyska}, D.~{Eisenstein}, S.~{Bose}, L.~H. {Garrison} and
  N.~{Maksimova}, \emph{{COMPASO: A new halo finder for competitive assignment
  to spherical overdensities}},
  \href{https://doi.org/10.1093/mnras/stab2980}{\emph{\mnras} {\bfseries 509}
  (2022) 501} [\href{https://arxiv.org/abs/2110.11408}{{\ttfamily
  2110.11408}}].

\bibitem{garrison16}
L.~H. {Garrison}, D.~J. {Eisenstein}, D.~{Ferrer}, M.~V. {Metchnik} and P.~A.
  {Pinto}, \emph{{Improving initial conditions for cosmological N-body
  simulations}}, \href{https://doi.org/10.1093/mnras/stw1594}{\emph{\mnras}
  {\bfseries 461} (2016) 4125}
  [\href{https://arxiv.org/abs/1605.02333}{{\ttfamily 1605.02333}}].

\bibitem{jing05}
Y.~P. {Jing}, \emph{{Correcting for the Alias Effect When Measuring the Power
  Spectrum Using a Fast Fourier Transform}},
  \href{https://doi.org/10.1086/427087}{\emph{\apj} {\bfseries 620} (2005) 559}
  [\href{https://arxiv.org/abs/astro-ph/0409240}{{\ttfamily
  astro-ph/0409240}}].

\bibitem{bel15}
J.~{Bel}, K.~{Hoffmann} and E.~{Gazta{\~n}aga}, \emph{{Non-local bias
  contribution to third-order galaxy correlations}},
  \href{https://doi.org/10.1093/mnras/stv1600}{\emph{\mnras} {\bfseries 453}
  (2015) 259} [\href{https://arxiv.org/abs/1504.02074}{{\ttfamily
  1504.02074}}].

\bibitem{zennaro21}
M.~{Zennaro}, R.~E. {Angulo}, M.~{Pellejero-Ib{\'a}{\~n}ez}, J.~{St{\"u}cker},
  S.~{Contreras} and G.~{Aric{\`o}}, \emph{{The BACCO simulation project:
  biased tracers in real space}}, {\emph{arXiv e-prints} (2021)
  arXiv:2101.12187} [\href{https://arxiv.org/abs/2101.12187}{{\ttfamily
  2101.12187}}].

\bibitem{yuan18}
S.~{Yuan}, D.~J. {Eisenstein} and L.~H. {Garrison}, \emph{{Exploring the
  squeezed three-point galaxy correlation function with generalized halo
  occupation distribution models}},
  \href{https://doi.org/10.1093/mnras/sty1089}{\emph{\mnras} {\bfseries 478}
  (2018) 2019} [\href{https://arxiv.org/abs/1802.10115}{{\ttfamily
  1802.10115}}].

\bibitem{hadzhiyska20}
B.~{Hadzhiyska}, S.~{Bose}, D.~{Eisenstein}, L.~{Hernquist} and D.~N.
  {Spergel}, \emph{{Limitations to the `basic' HOD model and beyond}},
  \href{https://doi.org/10.1093/mnras/staa623}{\emph{\mnras} {\bfseries 493}
  (2020) 5506} [\href{https://arxiv.org/abs/1911.02610}{{\ttfamily
  1911.02610}}].

\bibitem{hadzhiyska21}
B.~{Hadzhiyska}, S.~{Bose}, D.~{Eisenstein} and L.~{Hernquist},
  \emph{{Extensions to models of the galaxy-halo connection}},
  \href{https://doi.org/10.1093/mnras/staa3776}{\emph{\mnras} {\bfseries 501}
  (2021) 1603} [\href{https://arxiv.org/abs/2008.04913}{{\ttfamily
  2008.04913}}].

\end{thebibliography}\endgroup
\bibliographystyle{JHEP}

\end{document}